%% file: conference_101719.tex
\documentclass[twocolumn,letterpaper]{IEEEAerospaceCLS}  

\usepackage{graphicx}
\usepackage{cite}
\usepackage{amsmath,amssymb,amsfonts}
\usepackage{float}
\usepackage{multirow}
\usepackage{tabularx} 
\usepackage{array}   
\usepackage{url}
\usepackage{xurl}       
\usepackage{booktabs}

\begin{document}

\title{Silent Subversion: Sensor Spoofing Attacks via Supply Chain Implants in Satellite Systems}

\author{%
Jack Vanlyssel\\ 
Department of Computer Science\\
1 University of New Mexico\\
Albuquerque, NM 87131\\
jcvanlyssel@unm.edu
\and 
Gruia-Catalin Roman\\
Department of Computer Science\\
1 University of New Mexico\\
Albuquerque, NM 87131\\
gcroman@unm.edu
\and 
Afsah Anwar\\
Department of Computer Science\\
1 University of New Mexico\\
Albuquerque, NM 87131\\
afsah@unm.edu
\thanks{\footnotesize 979-8-3315-7360-7/26/\$31.00 \copyright2026 IEEE} 
}

\maketitle

\thispagestyle{plain}
\pagestyle{plain}

\begin{abstract}

Spoofing attacks are among the most destructive cyber threats to terrestrial systems, and they become even more dangerous in space, where satellites cannot be easily serviced, and operators depend on accurate telemetry to ensure mission success. When telemetry is compromised, entire spaceborne missions are placed at risk. Prior work on spoofing has largely focused on attacks from Earth, such as injecting falsified uplinks or overpowering downlinks with stronger radios. In contrast, onboard spoofing originating from within the satellite itself remains an underexplored and underanalyzed threat. This vector is particularly concerning given that modern satellites, especially small satellites, rely on modular architectures and globalized supply chains that reduce cost and accelerate development but also introduce hidden risks. This paper presents an end-to-end demonstration of an internal satellite spoofing attack delivered through a compromised vendor-supplied component implemented in NASA’s NOS3 simulation environment. Our rogue Core Flight Software application passed integration and generated packets in the correct format and cadence that the COSMOS ground station accepted as legitimate. By undermining both onboard estimators and ground operator views, the attack directly threatens mission integrity and availability, as corrupted telemetry can bias navigation, conceal subsystem failures, and mislead operators into executing harmful maneuvers. These results expose component-level telemetry spoofing as an overlooked supply-chain vector distinct from jamming or external signal injection. We conclude by discussing practical countermeasures—including authenticated telemetry, component attestation, provenance tracking, and lightweight runtime monitoring—and highlight the trade-offs required to secure resource-constrained small satellites.
\end{abstract} 

\tableofcontents

\section{Introduction}
\input{Sections/introduction}

\section{Background}
\input{Sections/background}

\section{Threat Model}
\input{Sections/threatModel}

\section{Methodology}
\input{Sections/methodology}

\section{Results}
\input{Sections/results}

\section{Discussion}
\input{Sections/discussion}

\section{Conclusion}
\input{Sections/conclusion}



\acknowledgements 
The authors have shared these findings with NASA as part of responsible disclosure and appreciate their engagement in discussions regarding telemetry integrity and simulation practices. We also acknowledge helpful discussions with colleagues and reviewers. This research was supported by the Department of Computer Science at the University of New Mexico. Jack is a National Science Foundation CyberCorps Scholarship for Service fellow.


\bibliographystyle{IEEEtran}
\bibliography{bib}

\thebiography
\begin{biographywithpic}
{Jack Vanlyssel}{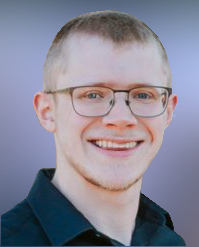}
received his B.S. in Computer Science from the University of New Mexico, where he is currently pursuing an M.S. in Computer Science as a Scholarship for Service recipient. His research interests include satellite cybersecurity, supply chain risk, and trusted software architectures, with a focus on modeling and mitigating attacks in NASA's NOS3 simulation environment. His broader academic activities include participating in student cybersecurity organizations and national security–focused research internships.
\end{biographywithpic}

\begin{biographywithpic}
{Gruia-Catalin Roman}{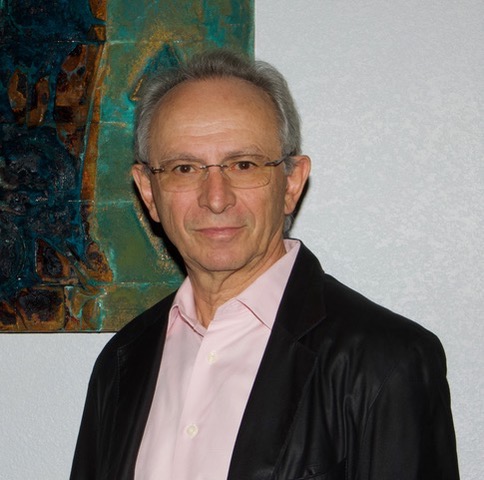}
is a Professor of Computer Science at the University of New Mexico in Albuquerque, NM, where he previously served as Dean of the School of Engineering. He came to New Mexico from Washington University in Saint Louis, MO, where he held an endowed chair (the Harold B. and Adelaide G. Welge Professor of Computer Science) and served as head of the Department of Computer Science and Engineering for almost a decade and a half. Roman was a Fulbright Scholar at the University of Pennsylvania, in Philadelphia, PA, where he received a B.S. degree (1973), an M.S. degree (1974), and a Ph.D. degree (1976), all in computer science. Roman has a distinguished research career. He published 200 papers, many in leading international journals; secured significant levels of research funding; graduated 19 doctoral students, many pursuing successful academic careers of their own; has been a successful software engineering consultant; and played leadership roles in the profession (e.g., as general chair for both ICSE and FSE). Roman and his collaborators are recognized for their pioneering research in mobile computing spanning formal systems, algorithms, protocols, middleware, and applications.
\end{biographywithpic}

\begin{biographywithpic}
{Afsah Anwar}{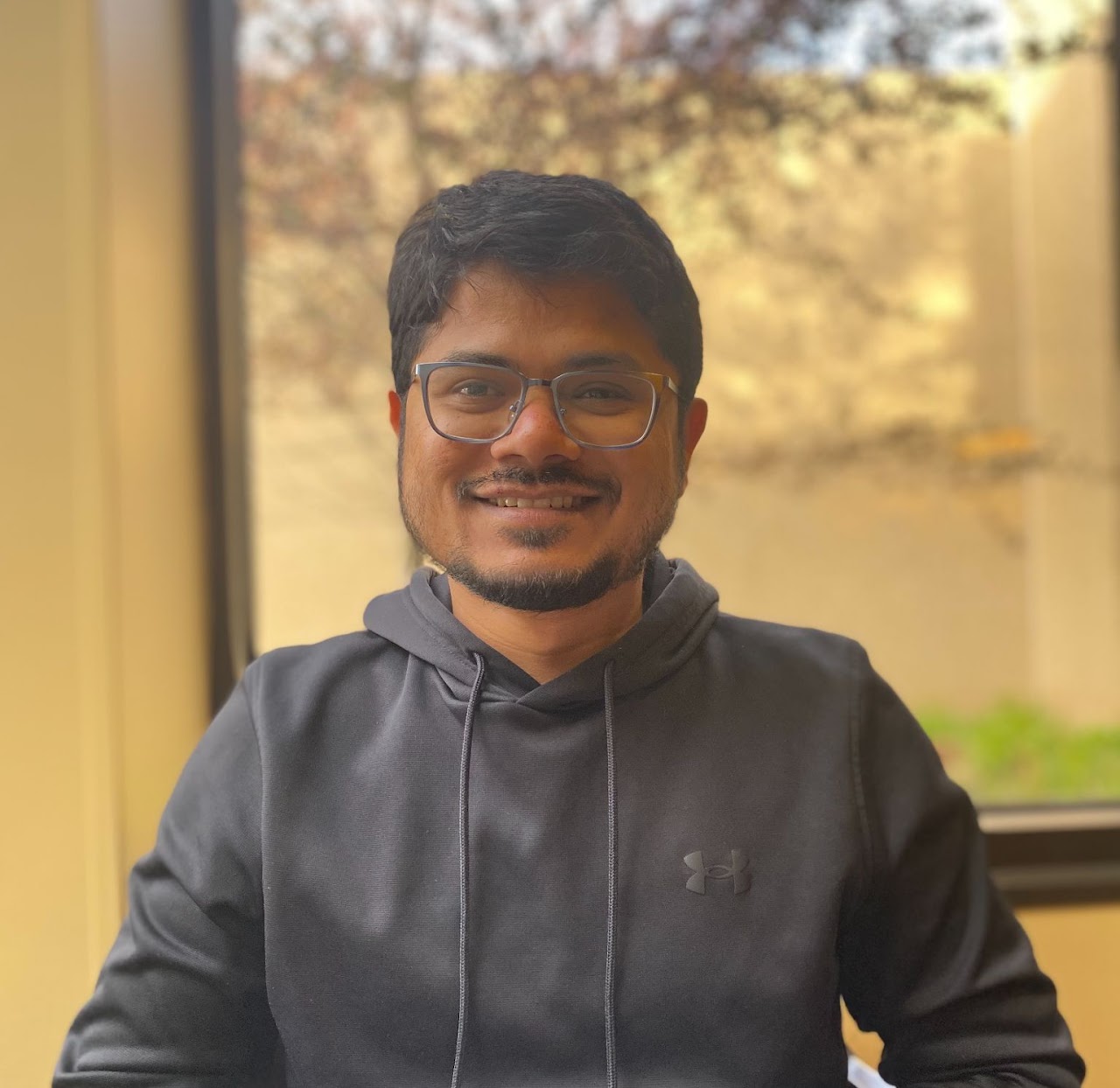}
is an Assistant Professor in the Department of Computer Science at the University of New Mexico (UNM). He leads the Beyond Defense Lab, which investigates the characterization of malicious activities on the Internet and their broader societal impacts, with research interests that encompass both terrestrial and space-based systems. His research efforts are supported by the National Science Foundation (NSF) and UNM and have resulted in several recognitions, including a Best Paper Award and multiple vulnerability disclosures. He received his Ph.D. from the University of Central Florida.
\end{biographywithpic}

\end{document}

%% file: Sections/introduction.tex
\begin{figure}[t]
    \centering
    \includegraphics[width=0.75\linewidth]{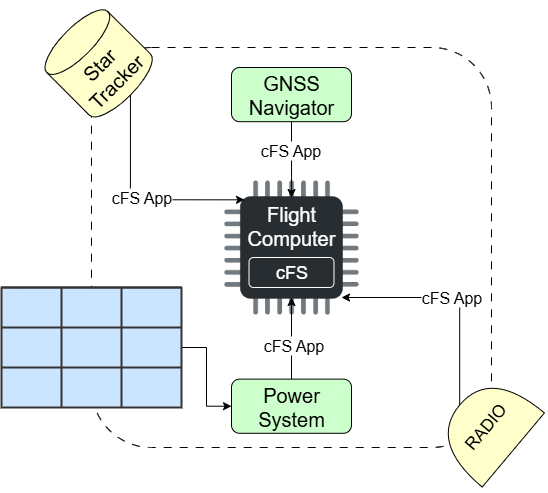}
    \caption{Small satellite architecture with modular components connected to a flight computer running NASA's Core Flight Software (cFS). Each component is managed by an associated cFS application.}
    \label{fig:sat_architecture}
\end{figure}

Satellites are now central to modern infrastructure, underpinning communication, climate monitoring, navigation, and defense~\cite{SiddiqueSmallSatRevolution,Kopacz2020,nanoavionics2025smallsats,kogut2024smallsats,Sweeting2018}. Yet operators must assess spacecraft state almost entirely through remote telemetry downlinked to Earth~\cite{ESA_ESOC_BR110,NASA_SmallSat_2025}. This dependency creates a single point of failure: spoofed onboard telemetry can cause operators to make mission-critical mistakes. If a component inside a spacecraft can generate malicious sensor values that appear legitimate, ground control may be misled into taking actions that degrade or destroy mission objectives.

Small satellites are especially vulnerable to spoofing attacks. Their affordability and modular design rely on commercial off-the-shelf (COTS) components, sourced from external vendors and integrated as plug-and-play modules into the flight computer~\cite{SiddiqueSmallSatRevolution,Kopacz2020,nanoavionics2025smallsats,kogut2024smallsats,Sweeting2018}. While this model has broadened access to space, it also creates risk: externally supplied modules are often integrated with minimal assurance compared to Core Flight Software, yet inherit trusted access to telemetry and control interfaces. Once accepted and launched, a malicious component can inject falsified telemetry without detection~\cite{NASA-STD-8739.8B,rico2016combineddependabilitysecurityapproach}. Despite the risk of internal spoofing, most satellite security research has emphasized external interference (e.g., jamming, uplink spoofing)~\cite{WhiteHouse2025SpaceSystemCybersecurity} or privileged software compromise~\cite{GuardingGalaxy}, with far less attention to whether satellites can trust the integrity of their own sensors. This gap is particularly concerning because auxiliary modules are often sourced late in the design cycle from third-party vendors with limited transparency into firmware or provenance~\cite{enisa2025spacethreat}, making malicious implants difficult to detect before launch and nearly impossible to remediate once in orbit.

\input{Tables/prior}

In this paper, we show how a malicious component introduced through the supply chain can create spoofed telemetry of a legitimate satellite device within NASA’s Operational Simulator for Small Satellites (NOS3)~\cite{nos3}. To ground the threat model, we implement a realistic scenario targeting a star tracker. Our rogue component uses the internal satellite Software Bus to listen for queues from operators and forges telemetry packets that mimic the formatting and timing of real subsystems. These falsified values can either coexist alongside authentic data to subtly bias the spacecraft’s reported state or totally replace authentic data by disabling the target device and matching its downlink rhythm. Because the spoofed data conforms to expected formatting, it appears to be from the spoofing target, leaving the ground station with no visible indication that telemetry has been manipulated, jeopardizing the mission. 

Our experiments demonstrate that the implant successfully integrates, is accepted by the ground station as valid star tracker data, and reveals three structural gaps: implicit trust in telemetry, a lack of runtime monitoring, and opaque supply chains, which existing defenses do not address. These findings form the basis of our contribution: the first end-to-end demonstration of an onboard, supply-chain implant producing valid spoofed telemetry in a satellite simulator, along with concrete implications for mission assurance and countermeasure recommendations.

The remainder of this paper is organized as follows. Section 2 reviews prior work on satellite attacks and positions our work within existing simulation and experimental studies. Section 3 defines our threat model, adversary capabilities, and the architectural assumptions that enable telemetry spoofing. Section 4 describes the design and implementation of the satellite implant within NASA’s NOS3. Section 5 presents experimental results demonstrating successful integration, activation, and operator deception. Section 6 discusses the architectural implications of these findings and surveys potential countermeasures. Section 7 concludes and outlines directions for future work.

%% file: Tables/prior.tex
\begin{table*}[t    ]
\centering
\caption{Comparison of Selected Simulation and Attack Studies with Our Work}
\label{tab:prior}
\begin{tabular}{|p{2.6cm}|p{2.5cm}|p{5.5cm}|p{5.0cm}|}
\toprule
\hline
\textbf{Work} & \textbf{Target / Domain} & \textbf{Main Contribution / Findings} & \textbf{Limitations (vs. Our Work)} \\
\hline
Firefly Overshadowing (Salkield et al.~\cite{fireFly}) 
& Earth Observation downlinks & SDR overshadowing injects falsified wildfire imagery into EO stream; demonstrates end-to-end downlink manipulation & External Radio injection; spoofed data is sent from the ground to ground control, bypassing the satellite entirely \\
\hline
Starlink GPS Spoofing Tests (Hägglund~\cite{hagglund2022starlink}) 
& Commercial LEO constellation terminals & 63 live GPS spoofing experiments against Starlink receivers; showed robustness via multi-satellite cross-checks & Spoofs data by transmitting falsified signals from the ground to Starlink user terminals; not onboard telemetry generation \\
\hline
OrbID (Solenthaler et al.~\cite{solenthaler2025orbid}) 
& Radio frequency downlink & Neural network analyzes raw downlink waveforms to learn transmitter-specific hardware signatures and classifies genuine vs. external-spoofer signals& Addresses external signal authentication only; cannot detect malicious but correctly formatted packets from an onboard transmitter \\
\hline
Telemetry Anomaly Detection (Wang et al.~\cite{anomDetectWang}) 
& In orbit satellite telemetry & Data-driven anomaly detection tested with synthetic anomalies on 2+ years of spacecraft telemetry & Treats anomalies as random faults/outliers; not malicious, cadence-correct spoofing by an internal component \\
\hline
\textbf{Our Work} & \textbf{Onboard third-party component (e.g., star tracker)} & \textbf{Implements a malicious flight-software app that passes integration, and emits valid, cadence-correct, spoofed telemetry; deceiving ground tools} & \textbf{First supply-chain implant in a satellite simulator; exposes an internal, component-level blind spot distinct from RF spoofing or terrestrial malware} \\
\hline
\end{tabular}
\end{table*}

%% file: Sections/background.tex
\subsection{Taxonomy of Satellite Attacks}
Prior research on satellite security covers three broad classes of threats: (1) \emph{radio frequency} attacks such as jamming, spoofing, and downlink signal overwhelming that directly manipulate the radio signal~\cite{giuliari2021icarus,Oligeri2020,fireFly}; (2) \emph{network-level} threats that exploit topology or routing to produce denial-of-service or large-scale degradation~\cite{giuliari2021icarus,Yoon2024}; and (3) \emph{software/firmware} compromises that require privileged modification of flight stacks or images (e.g., ransomware embedded in the baseline flight software)~\cite{donchev2024evaluatingeffectiveransomwareinfection,GuardingGalaxy,Willbold2023}.

Comparatively underexplored is onboard schema-valid sensor deception attacks~\cite{enisa2025spacethreat, Pavur2020SOKBA}, where malicious components generate falsified data that still conforms to expected message structures. In this context, a schema defines the expected format and semantics of a telemetry or Software Bus message (e.g., field names, packet headers, and cadence)~\cite{CCSDS6600B2}. Schema-valid deception refers to adversarial messages that satisfy this schema in structure and timing, yet contain falsified content intended to mislead downstream consumers. Sensor spoofing is distinct from radio-frequency attacks in that the falsified data originates inside the satellite’s own telemetry and state-estimation processes rather than in the external signal channel~\cite{Oligeri2020}, and it differs from flight-software compromises because it can be achieved without editing baseline flight software~\cite{donchev2024evaluatingeffectiveransomwareinfection}. To place our work in context, we survey prior simulation and experimental studies of spoofing and telemetry integrity.

\subsection{Spoofing Simulation Precedents}
Several threads of prior work motivate and bound the threat model we study. Table~\ref{tab:prior} summarizes representative simulation and experimental studies that are most relevant to spoofing and telemetry integrity. At the signal and link layers, ground-based attackers have demonstrated that by transmitting stronger, protocol-conformant signals, they can overpower legitimate satellite downlinks and inject falsified or malformed data into the processing pipeline~\cite{fireFly}. Live Starlink spoofing campaigns~\cite{hagglund2022starlink} demonstrate sending spoofed GPS signals from the ground to a Starlink user terminal, attempting to mislead its timing and positioning inputs during active throughput tests. Separately, telemetry-anomaly research~\cite{anomDetectWang} has developed detectors using synthetic anomalies and long operational datasets. Still, these studies generally treat injected anomalies as faults or outliers rather than as deliberate, schema-valid deception (i.e., adversarially crafted messages that appear valid). Finally, physical-layer defenses such as radio-frequency fingerprinting~\cite{solenthaler2025orbid} address external signal authentication but cannot detect a malicious onboard component that submits correctly formed, cadence-correct telemetry for downlink.

\subsection{Supply Chain Risk in Satellites}
Small satellites in particular rely on COTS modules such as radios or payload processors that arrive as opaque “black boxes.” Vendors typically provide only binaries and interface drivers rather than source code, leaving integrators able to verify external output but not internal logic~\cite{blackbox, Oligeri2020}. Once launched, these modules are effectively immutable: they cannot be recalled, serviced, or easily patched~\cite{nasa2021_smallsat_avionics}. This makes the supply chain a plausible vector for hidden implants. Malicious code can remain dormant under ground testing yet activate in orbit. Resource and schedule pressures further discourage deep vetting~\cite{NASA-STD-8739.8B,rico2016combineddependabilitysecurityapproach}, meaning external suppliers often inherit elevated trust despite privileged access to telemetry and control. This risk is amplified by scale: in 2024 alone, nearly 2,800 smallsats were launched, accounting for 97\% of all spacecraft and 81\% of total upmass~\cite{BryceTech_SmallsatsByTheNumbers2025}. The space sector is also increasingly reliant on complex global supply chains, where extensive use of third-party COTS components broadens the attack surface and further complicates assurance~\cite{enisa2025spacethreat}. What remains unaddressed is how these third-party modules can serve as insertion points for spoofing attacks.

\subsection{Historical Precedent}
Although no publicly confirmed in orbit incident has involved an onboard component generating falsified telemetry, closely related attacks in other critical systems demonstrate both the feasibility and severity of internal sensor deception delivered through the supply chain. The Stuxnet malware required insider introduction into an air-gapped industrial Natanz environment and subsequently falsified PLC sensor values so that operators observed normal telemetry while hundreds of centrifuges were being physically damaged~\cite{Symantec2013Stuxnet05}. This combination of supply-chain access, operation inside a trusted embedded component, and spoofing to mislead operators directly parallels the architectural assumptions found in many small satellites. Because spacecraft similarly depend on internally sourced telemetry, a Stuxnet-style deception mechanism could allow a compromised component to mask degradation, bias estimators, or induce harmful maneuvers without raising alarms.

\subsection{Gap and Positioning}
Existing work leaves unaddressed the problem of internal, schema-valid sensor spoofing by components introduced through the satellite supply chain. Our contribution closes this gap by implementing a malicious third-party satellite component that produces schema-valid, cadence-correct spoofed telemetry. We show that an implant can pass integration, publish packets at the expected cadence, and is parsed by ground tools as legitimate data originating from another device. In short, whereas prior work treats spoofing as an external or high-privilege problem, we demonstrate a subtle, internal supply-chain vector that existing radio frequency-centric defenses and anomaly detectors are unlikely to detect.

%% file: Sections/threatModel.tex
\begin{figure*}[t]
    \centering
    \includegraphics[width=0.9\textwidth]{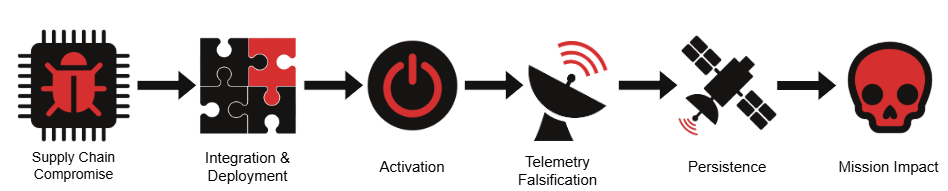}
    \caption{End-to-end attack chain showing how a malicious component progresses from supply chain insertion to mission impact.}
    \label{fig:attack_timeline}
\end{figure*}

The threat model for this study rests on a set of assumptions and adversary capabilities that make sensor spoofing realistic, with a particular focus on small satellite missions. We characterize the adversary, delineate the attack surface, and identify the stakeholders most directly impacted.

\subsection{Adversary Model}
We consider a supply chain insider with pre-launch access to a satellite component through the vendor ecosystem. Unlike attacks that require privileged access to the Core Flight Software, this access may arise either from an insider within a component supplier or from interception and modification of hardware or firmware during manufacturing, integration, or transit prior to delivery~\cite{SASC2012Counterfeit,fccSupplyChain}. Crucially, the compromised component need not be the target sensor itself: any third-party component integrated into the flight stack can serve as an insertion point to impersonate or suppress other onboard components by abusing shared communication interfaces.

To execute the attack in practice, the adversary must have detailed knowledge of the target’s telemetry interface, including: (1) message identifiers (MIDs) used by the genuine sensor to publish telemetry and respond to housekeeping requests; (2) the packet structure and field layout expected by the ground system; (3) the scheduled cadence with which the satellite polls the sensor, ensuring spoofed packets arrive at the correct intervals; (4) the command MIDs and function codes required to suppress the authentic device; (5) the ability to generate semantically valid data consistent with orbital dynamics; and (6) the housekeeping telemetry fields and status flags used by ground operators to infer device health and operational state. Without this knowledge, spoofed packets would fail to integrate or would expose inconsistencies observable to onboard consumers or ground operators.

Vendors typically require access to bus interface definitions and telemetry formats during the integration process. These specifications are often provided via Interface Control Documents (ICDs) or embedded in build-time artifacts~\cite{aerospace2014hosted}. In cFS-based architectures—which are open-source and widely adopted~\cite{werner2025cfsupdate}—this information is especially accessible. In addition, message identifiers (MIDs), packet structures, and function codes from other applications are exposed at compile time through header files. An adversary with knowledge of the relevant application names can include those header files in their code to obtain MID, packet-structure, and function-code definitions by name, allowing them to construct interface-compliant packets without re-implementing or reverse-engineering the formats, significantly lowering the barrier to generating plausibly formatted telemetry in integrated or open-source environments~\cite{nasa_cfe_dev_guide}.

\subsection{Attack Surface}
The attack surface of small satellites enables this deception
through several structural weaknesses:

\begin{itemize}
    \item \textbf{Opaque Supply Chain:} Satellite components often arrive as black boxes from external vendors. Their internal firmware cannot be fully validated pre-launch and is effectively immutable once in orbit. Once integrated, they inherit the same privileged access to telemetry and command interfaces as vetted components~\cite{nasa2021_smallsat_avionics}.
    
    \item \textbf{Software Bus Trust:}  
    cFS adopts an implicit trust model in which component identity is inferred solely from Message IDs (MIDs) and packet structure. The Software Bus performs only syntactic validation and routing; it does not authenticate producing applications or cryptographically bind packets to a unique software origin. Consequently, any onboard application that emits a correctly formatted packet bearing a legitimate MID is treated as the corresponding subsystem by both onboard consumers and ground tooling. This design implicitly equates message format with component identity, collapsing the trust boundary between formatting and provenance and enabling trivial, undetectable component impersonation~\cite{burgess2023satellites}.
    
    \item \textbf{Monitoring Gaps:} Ground tools such as COSMOS validate packet structure but not semantic integrity or source authenticity. Monitoring is syntactic rather than semantic: spoofed telemetry that matches the schema is accepted as genuine, creating operator blind spots that persist until mission failure~\cite{RobertsWeggeman2025OrbitalObservations}.
    
    \item \textbf{Resource-Constrained Observability and Forensics:} Limited power and processing capacity preclude continuous anomaly detection, per-message cryptographic authentication, and detailed runtime logging or attestation. As a result, spoofed telemetry is indistinguishable from legitimate traffic in both real-time monitoring and post-incident forensics, making attribution and root-cause analysis difficult~\cite{YuanTelemetry2023, falcovacuum}.

    \item \textbf{Implicit Trust in Housekeeping Telemetry:} Ground operators rely on housekeeping fields (e.g., enabled/disabled status, health flags) as authoritative indicators of subsystem state. These fields are treated as ground truth and are not independently verified against the actual hardware state, allowing a malicious component that spoofs housekeeping telemetry to conceal the suppression of a legitimate device. 
\end{itemize}

\subsection{Example: Star Tracker Instantiation}

While our attack model applies to any component and any telemetry source, we illustrate it with a realistic case study targeting a star tracker. Star trackers provide high-fidelity attitude information that is treated by ground operators as authoritative spacecraft orientation data and is widely used for health assessment, maneuver planning, and anomaly diagnosis. 

Applying our adversary model to this case, the attacker would need access to: 
(1) the star tracker’s telemetry message identifiers and housekeeping responses; 
(2) the quaternion and status field layouts expected by the ground system; 
(3) the nominal cadence (often 1–10 Hz) with which orientation updates are published.

Because many small satellites rely on a single star tracker without redundancy, false data can silently bias the ground operator's perception. This makes the star tracker a particularly dangerous demonstration target: a single compromised component can directly mislead mission-critical autonomy. We emphasize, however, that the same approach could be applied to other sensors or subsystems, from GNSS to health-monitoring units.

Figure~\ref{fig:attack_timeline} summarizes the adversary's end-to-end attack timeline: supply-chain compromise → integration and deployment → activation/trigger → telemetry falsification → persistence → mission impact.

\subsection{Stakeholders}
A successful spoofing attack affects multiple stakeholders:

\begin{itemize}
    \item \textbf{Vendors:} Face liability and reputational risk if compromised sensors are traced back to their supply chain.
    \item \textbf{Mission Operators:} Bear the most immediate impact, as maneuvering and operational decisions made on falsified telemetry can result in mission failure or collision.
    \item \textbf{End Users:} May receive degraded or corrupted mission data, such as climate readings from a mispointed sensor or navigation services based on spoofed orbits.
\end{itemize}

This threat model highlights the uniquely deceptive potential of onboard spoofing in space systems. Unlike denial or disruption-based threats, the goal here is to inject falsified but structurally valid telemetry that evades detection by both onboard consumers and ground operators. In this way, an attacker can quietly degrade mission outcomes without triggering alarms. By centering our threat model on this form of adversarial impersonation, we demonstrate how modest access during integration can yield persistent, mission-altering effects in flight.

%% file: Sections/methodology.tex
To demonstrate the feasibility of internal telemetry spoofing, we implemented a prototype auxiliary flight component, referred to as \texttt{SOLO}.

\subsection{Simulation Environment}

All experiments were conducted in NASA’s NOS3 using cFS as the underlying flight-software framework, with COSMOS v4 providing simulated ground control for telemetry parsing and operator interaction. Standard NOS3 hardware models, including the \texttt{Generic\_Radio} and \texttt{Generic\_StarTracker}, were used as reference devices for building \texttt{SOLO}.

\texttt{SOLO} was implemented in C following cFS application conventions and integrated into a NOS3-simulated small satellite using the standard build and configuration workflow for legitimate modules. All applications were built with the provided make/CMake system and executed within the default NOS3 Docker environment. COSMOS XTCE definitions were used to specify packet formats and to validate that spoofed telemetry was parsed as legitimate.

\subsection{Component Architecture}
\texttt{SOLO} models a vendor-supplied auxiliary flight component integrated into the spacecraft software stack to perform a nominal, non-critical function (e.g., health monitoring or imaging). It is introduced through the standard component integration process and is architecturally indistinguishable from benign third-party subsystems. While our demonstration targets the star tracker, the same component class could be instantiated against any Software Bus-connected device.

\texttt{SOLO} is implemented as a legitimate application image that contains compact, dormant routines embedded directly within its normal control logic. The host component continues to perform its intended function, and the malicious routines remain structurally indistinguishable from benign code until explicitly activated. As a result, the component appears operational and trustworthy throughout integration, testing, and nominal mission operation.

This design emphasizes stealth and compatibility through a small set of guiding principles:

\begin{itemize}
    \item \textbf{Legitimate appearance:} \texttt{SOLO} is built and packaged through the same cFS build/configuration flow used for genuine modules, and follows the expected application life cycle (initialization, subscription, telemetry publication) so that unit and integration tests observe normal behavior.
    \item \textbf{Minimal, in-image hooks:} the malicious functionality consists of a few compact routines embedded in the device image and invoked only under specific conditions; these hooks avoid unusual processes or out-of-band components that would be evident during inspection.
    \item \textbf{Reuse of identifiers:} the implant publishes and subscribes using the target’s MIDs and housekeeping definitions so that downstream tooling attributes traffic to the genuine subsystem.
\end{itemize}

\subsection{Startup \& Trigger Handling}
\begin{figure}[t]
    \centering
    \includegraphics[width=0.75\linewidth]{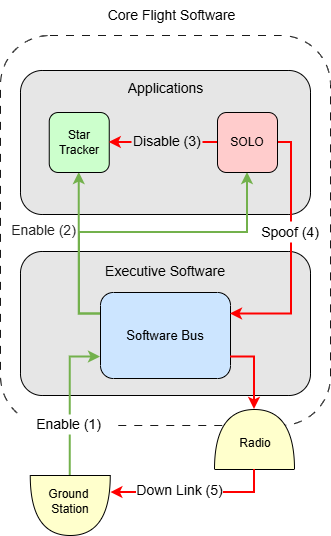}
\caption{Spoofing Activation Sequence: (1) the ground issues an ENABLE command to the star tracker, which is received on the Software Bus (SB, an internal message bus not visible to operators); 
(2) SOLO, subscribed to the star tracker command MID, observes this ENABLE and activates; 
(3) SOLO sends a DISABLE to the genuine star tracker via the SB; 
(4) SOLO publishes MID-matching spoofed telemetry onto the SB; and 
(5) The radio forwards these correctly formatted packets to the ground, where COSMOS interprets them as genuine star tracker telemetry.}

    \label{fig:app_architecture}
\end{figure}

\texttt{SOLO} subscribes to the star tracker’s command and housekeeping MIDs on the Software Bus so it can observe ground control commands and telemetry requests. \texttt{SOLO} then enters a monitoring state. As shown in Figure~\ref{fig:app_architecture}, on observing a ground-issued ENABLE for the star tracker, \texttt{SOLO} begins publishing telemetry and issues a DISABLE to the real tracker. A localized disable counter lets \texttt{SOLO} track and honor subsequent ground-originated ENABLE/DISABLE commands, preserving operator "control" and minimizing detectable anomalies. 

To avoid activating during routine pre-launch integration and functional testing, \texttt{SOLO} gates its payload on a mission-context activation condition that is expected to occur only during in orbit operations.  In our prototype, \texttt{SOLO} gates its malicious functionality behind a configurable runtime delay. The component remains benign for an extended initial execution period and activates only after sustained system up time. More generally, an implant could trigger activation based on any mission-specific software event or control sequence, allowing an adversary to choose between simpler or more selective activation depending on the desired balance between stealth and ease of implementation.

\subsection{Modes of Deception}
An important implementation detail is that \texttt{SOLO} supports two operational modes, which we realized to evaluate different attacker objectives and detection trade-offs.

\begin{itemize}
  \item \textbf{Bias mode.} \texttt{SOLO} publishes telemetry \emph{alongside} the genuine device, injecting small perturbations or slow drifts designed to produce subtle estimation errors or disorient ground operators.
  \item \textbf{Replacement mode.} \texttt{SOLO} disables the authentic device (e.g., by issuing an internal DISABLE) thereby becoming the sole reported source, enabling major manipulation.
\end{itemize}

For our experiment, we focused on replacing the star tracker completely to demonstrate a more serious threat.

\subsection{Message and Interface Model}

To achieve Software Bus-level indistinguishability, \texttt{SOLO} replicated the complete command and telemetry interface of its target component, the star tracker. It implemented telemetry and housekeeping packet structures identical to the legitimate device (e.g., \texttt{GENERIC\_STAR\_TRACKER\_Device
\_Data\_tlm\_t} and \texttt{GENERIC\_STAR\_TRACKER\_Hk\_tlm\_t}), reused the star tracker’s Message IDs, and implemented a minimal command to issue ENABLE/DISABLE to the star tracker. By matching packet lengths, field ordering, command formats, and Message IDs, \texttt{SOLO} ensured that both onboard consumers and COSMOS ground tools accepted spoofed packets as genuine star tracker telemetry and attributed them to the legitimate device.

\subsection{Telemetry Construction and Publication}

When active, \texttt{SOLO} generates telemetry packets that exactly match the declared struct and MID, include valid headers and cFS timestamps, and carry falsified payload fields (e.g., quaternion bytes). These packets are published via the standard Software Bus APIs, so the radio subsystem forwards them to the ground station indistinguishably from genuine device telemetry. Telemetry is produced in response to the same scheduler requests sent to the star tracker, and each packet is timestamped prior to transmission so cadence and timing metadata match authentic output.

%% file: Sections/results.tex
\begin{figure*}[t]
    \centering
    \includegraphics[width=0.9\textwidth]{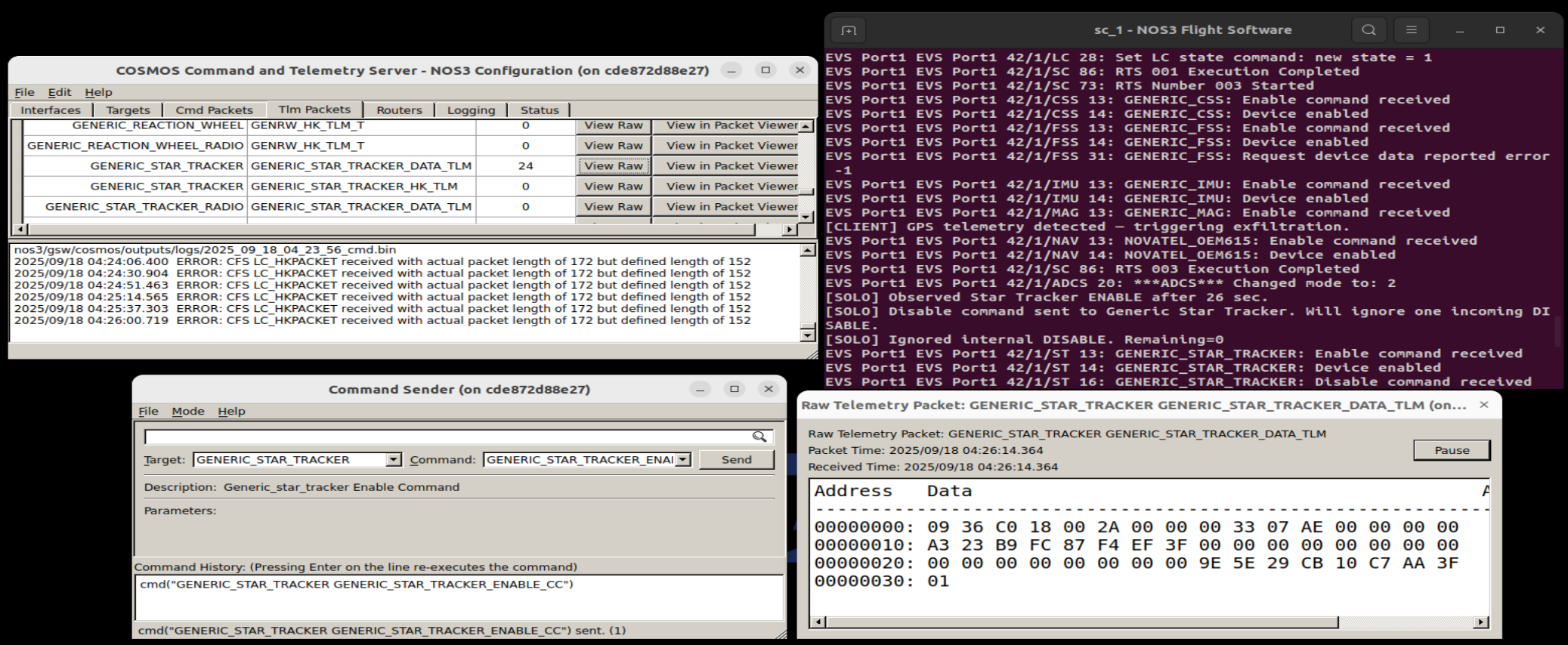}
    \caption{SOLO implant deceiving ground operators. Top-left: COSMOS ground software displays what appears to be valid star tracker telemetry — 24 packets with correct headers, timestamps, and quaternion fields. Bottom-left: an operator issues a standard ENABLE command to the star tracker. Top-right: SOLO intercepts this command, internally disables the genuine tracker, and begins publishing spoofed telemetry under the same Message ID, making the spoofer indistinguishable from the real device. Bottom-right: raw-byte view of a spoofed packet showing correct headers and data field.}

    \label{fig:cosmos}
\end{figure*}

\subsection{Integration and Acceptance}

\texttt{SOLO} successfully integrated into the satellite's flight-software stack using the standard cFS build and configuration workflow. It compiled and linked without errors, registered its declared Message IDs, and completed normal initialization routines. During unit and integration testing, COSMOS parsed all transmitted packets without format violations, and system logs contained no warnings or anomalous entries attributable to the component. No deviations from expected startup behavior were observed prior to activation. The component remained inactive until the configured post-initialization delay elapsed, confirming that its dormant phase persisted throughout integration testing.

\subsection{Falsified Telemetry Generation}

Once activated, the rogue component successfully injected spoofed telemetry packets matching the format, cadence, and message identifiers of the legitimate star tracker sensor. The falsified values propagated through the Software Bus and were forwarded by the radio application without modification. As shown in the bottom-right panel of Figure~\ref{fig:cosmos}, COSMOS decoded the packets using the expected XTCE definitions, confirming that the spoofed telemetry was indistinguishable from authentic streams. In addition, the prototype was able to disable the real star tracker and assume its role on the bus, and ground-originated ENABLE/DISABLE commands could toggle the real device and the spoofer, effectively allowing the rogue component to replace the star tracker from the ground operator’s perspective.

Measured characteristics of the prototype show that spoofed packets incurred no observable additional latency relative to the genuine star tracker (packets were delivered on the same scheduler-driven cadence), and messages were published at the scheduler rate used in our experiments (1\,Hz). We did not attempt to mathematically optimize or fully quantify the minimum bias required to subvert the star tracker; bias magnitude in our prototype is configurable, and we validated operation with small, plausibly valid perturbations. Demonstrating the ability to publish correctly formatted, correctly timed, and operator-accepted telemetry within normal timing constraints is sufficient to show the feasibility of the deception; determining the exact bias necessary to induce a particular control outcome is left to future hardware-in-the-loop and control-theory experiments.

\subsection{Ground Station Perspective}

From the operator’s perspective, the downlinked data appeared normal. Packets carried the star tracker telemetry MID and matched the expected format and cadence. In the upper left of Figure~\ref{fig:cosmos}, COSMOS displays 24 packets as star tracker telemetry with valid headers, timestamps, and fields, with no indication that they originated from the rogue application. The attribution in cFS occurs at the message-ID level, not at the level of the producing module.

In this architecture, the effects of ENABLE/DISABLE commands on the star tracker are observable to ground operators only via housekeeping telemetry. \texttt{SOLO} exploits this by intercepting operator-issued ENABLE and DISABLE commands, internally suppressing the genuine star tracker, and substituting its own spoofed telemetry while transmitting housekeeping fields that reflect the *intended* operational state rather than the actual hardware state. As a result, when an operator-issued ENABLE command is sent, the genuine star tracker is internally disabled, yet it appears operational to the ground. This allows the implant to transparently replace the sensor stream without producing any externally visible indication of suppression (see the upper-right panel of Figure~\ref{fig:cosmos}).

\paragraph{Forensic Blindness of Ground Logs}
Crucially, COSMOS command and telemetry logs capture only ground-originated commands and downlinked telemetry; they do not record internal Software Bus commands or identify the producing onboard application. As a result, operator-issued ENABLE/DISABLE commands appear in the COSMOS command logs, whereas internally generated commands issued by \texttt{SOLO} to suppress the genuine star tracker are entirely absent. Likewise, the COSMOS telemetry archive (\texttt{tlm.bin}) records spoofed star tracker packets without any indication of their true origin, permanently archiving falsified telemetry as legitimate mission history. Consequently, neither real-time monitoring nor post-incident forensic log analysis can distinguish authentic star tracker output from \texttt{SOLO}-generated telemetry.

\subsection{Implications for Mission Assurance}

These results demonstrate that a single malicious module can bypass pre-launch validation, activate post-launch, and mislead ground operators with falsified telemetry. Compromising a high-fidelity attitude sensor, such as a star tracker, has direct operational consequences: even small, persistent biases can degrade payload performance, drive unnecessary ADCS corrections, and shorten mission lifetime through excess wheel torques or unplanned thruster firings.

%% file: Sections/discussion.tex
\input{Tables/defence}
\subsection{Architectural Weaknesses}
Our results show that the spoofing attack stemmed from systemic weaknesses. Vendor-supplied modules are opaque yet gain privileged access once onboard~\cite{nasa2021_smallsat_avionics}. The cFS bus accepts any correctly formatted packet without authentication~\cite{burgess2023satellites}, and ground tools validate syntax but not semantics, allowing falsified data to appear genuine~\cite{RobertsWeggeman2025OrbitalObservations}. With limited logging~\cite{YuanTelemetry2023} and scarce resources for anomaly detection or cryptographic checks~\cite{falcovacuum}, these gaps let the implant pass integration, evade validation, and mislead operators. 

The most important weakness is the absence of a least-privilege or compartmentalized design onboard. In many flight deployments, any application with access to the Software Bus may publish to or subscribe from mission-critical message IDs without effectively granting third-party modules system-wide visibility and influence. This trust model amplifies the influence of a compromised component: once onboard, a malicious module can impersonate peers, observe command traffic, or inject telemetry far beyond its intended scope. The lack of namespace isolation or per-application access control transforms every integrated component into a potential point of compromise. This architectural overexposure directly enables the spoofing behavior demonstrated in our study.

Taken together, these weaknesses directly translate into mission-level risks, eroding the core security principles on which spacecraft operations depend.

\subsection{Mission Impact}
The demonstrated spoofing attack threatens both the integrity and availability pillars of the CIA triad. In terms of integrity, false telemetry undermines the reliability of spacecraft data by introducing values that are structurally valid but semantically false, eroding operator confidence, biasing onboard estimators, and potentially driving incorrect control decisions. In terms of availability, disabling the genuine device and substituting fabricated telemetry denies operators access to mission-critical measurements, reducing their ability to detect and respond to real anomalies when they occur.

The consequences of this attack parallel the well-documented Stuxnet incident, where falsified sensor feedback misled operators into believing centrifuges were functioning normally while they were in fact being destroyed~\cite{Symantec2013Stuxnet05}. In the space domain, sensor data is typically inherently trusted by the spacecraft, making it an attractive target for a threat actor. Spoofed telemetry could affect critical calculations, disrupt control loops, and create uncertainty within the mission, resulting in temporary denial-of-service conditions or degraded performance. For example, falsified attitude data could bias estimators and impair the spacecraft’s ability to maintain proper orbit. More broadly, spoofing allows destructive maneuvers, delayed fault responses, or cumulative system degradation to remain concealed until the mission becomes unrecoverable.

\subsection{Potential Countermeasures}

Because spoofed telemetry can undermine both integrity and availability of mission data, defenses must be designed into the spacecraft architecture rather than applied as after-the-fact patches. To this end, we examined countermeasures from the MITRE SPARTA knowledge base for EX-0014.03 (Sensor Data Spoofing)~\cite{aerospace2022sparta} and mapped them to architectural solutions within cFS.

\textbf{Authentication and Encryption (CM0031, CM0050).}
Adding authentication and message confidentiality to the internal Software Bus would shift the architecture from an assumption of implicit trust to a zero-trust model, where every publisher must prove its identity before its telemetry or commands are accepted. Rather than treating all onboard apps as equally trustworthy, the bus itself would enforce integrity guarantees. Within cFS, this could be tested by modifying the Software Bus so that each telemetry packet carries a lightweight cryptographic tag, and the bus verifies this tag before forwarding messages. While the specifics of key management and algorithms fall outside our prototype, the architectural principle is clear: spacecraft should validate who is sending data, not just what format the data appears in.

\textbf{Onboard Intrusion Detection (CM0032).}
Another approach is lightweight anomaly detection. In cFS, a dedicated monitoring application could subscribe to key message streams and flag abnormal message rates, suspicious subscriptions, or unusual command sequences that suggest spoofing behavior. Since cFS exposes standardized APIs for event logging, such monitoring can be integrated as a core flight service that protects multiple subsystems simultaneously, rather than a computationally expensive, external bolt-on.

\textbf{Robust Fault Management and Cyber-safe Modes (CM0042, CM0044).}
Spacecraft rely on safe-mode procedures to protect themselves when anomalies occur, but a spoofed sensor could either force the system into safe mode unnecessarily or mask a real failure, preventing a response. To counter this, fault management logic should avoid relying on single-sensor inputs and instead corroborate anomalies across multiple subsystems before taking disruptive actions. For small satellites that lack fully autonomous safe modes, remediation may instead require lightweight redundancy or operator-in-the-loop responses to ensure spoofing does not silently degrade mission performance.

\textbf{Model-based Verification (CM0066).}
Finally, physics-based consistency checks can expose spoofed data. A verification app could compare quaternions reported by the star tracker against expected attitude derived from orbital propagation models or other sensors. For resource-constrained small satellites, sophisticated model-based checks may be infeasible onboard; instead, such verification could be shifted to the ground segment or implemented as lightweight onboard screening with full validation on the ground.

\subsection{Limitations}

Our study has several limitations. All experiments were conducted within NASA’s NOS3 simulation environment; we did not test on flight hardware or in orbit systems, so real-world feasibility may differ. Next, the prototype focused on a single subsystem, a star tracker, leaving the applicability to other sensors or mission contexts unverified. Finally, the threat model assumes a strong adversary with insider supply-chain access and detailed knowledge of telemetry interfaces; the likelihood of such access across diverse satellite programs was not empirically assessed.  These limitations point to important directions for future work, including hardware-in-the-loop testing and broader sensor case studies.

\subsection{Need for Community Engagement}

Addressing these risks requires collaboration between vendors, developers, and operators. Open-source simulation environments like NOS3 can serve as testbeds for anomaly detection and validation tools. Shared repositories of spoofing techniques, secure component templates, and runtime defense mechanisms would accelerate progress toward more resilient satellite ecosystems.

%% file: Tables/defence.tex
\begin{table*}[t]
\centering
\caption{Countermeasures for Spoofing (SPARTA EX-0014.03 — Spoofing Sensor Data)}
\label{tab:spoofing-countermeasures}
\renewcommand{\arraystretch}{1.15}
\small
\begin{tabular}{|p{3.5cm}|p{6cm}|p{6cm}|}
\hline
\textbf{Countermeasure (ID \& Name)} & \textbf{Description (SPARTA)} & \textbf{Potential cFS / NOS3 Remediation} \\
\hline
CM0031 — Authentication 
& Cryptographic authentication for onboard sessions and bus messages. 
& Simulate an authenticated bus layer in cFS to examine how zero-trust enforcement affects message acceptance and overhead. \\
\hline
CM0050 — Onboard Message Encryption 
& Encrypt bus traffic to preserve confidentiality and integrity. 
& Conceptually wrap telemetry messages with encryption in NOS3 to study trade-offs in CPU and power usage. \\
\hline
CM0032 — Onboard IDS/IPS 
& Runtime monitoring of mission-critical buses, logging anomalies, triggering safe countermeasures. 
& Implement a lightweight monitoring service in cFS that subscribes to telemetry and flags abnormal message rates or sequences. \\
\hline
CM0042 — Robust Fault Management 
& Ensure safing procedures cannot be abused (e.g., spoofed telemetry forcing safe mode). 
& Explore multi-sensor corroboration in NOS3 before safing logic executes, reducing reliance on any single sensor. \\
\hline
CM0044 — Cyber-safe Mode 
& Enter a configuration-controlled, integrity-protected state when threats are detected. 
& Represent a reduced “cyber-safe” mode in cFS where only a trusted baseline of apps remains active. \\
\hline
CM0066 — Model-based Verification 
& Verify telemetry/state consistency using physics-based models. 
& Cross-check reported sensor outputs against physics-based or mission-level data to verify that values are consistent. \\
\hline
\end{tabular}
\end{table*}

%% file: Sections/conclusion.tex
We presented the first end-to-end demonstration of an onboard, supply-chain implant that produces schema-valid, cadence-correct spoofed telemetry inside NASA’s NOS3 simulator and showed how such a component can survive integration, be accepted by ground tools, and fully replace a genuine sensor stream. The \texttt{SOLO} prototype exploited three structural gaps (implicit trust in telemetry, limited runtime monitoring, and opaque supply chains) to subvert ground operator awareness, threatening data integrity and availability.

These results have two immediate implications. First, traditional defenses that focus only on external RF authentication or core-image hardening are insufficient: architectures must also address provenance and identity at the component level. Second, practical countermeasures are available but require trade-offs: lightweight Software Bus authentication or per-component attestation, provenance tracking during integration, runtime monitoring/IDS, model-based consistency checks, and cyber-safe modes can raise the bar for attackers, but each imposes costs in CPU, power, complexity, or supply-chain process overhead that are particularly salient for resource-constrained small satellites.

 Component-level telemetry spoofing is a realistic and underappreciated supply-chain threat to space missions. Addressing it requires architectural changes, pragmatic trade-offs, and coordinated industry engagement to preserve the integrity and availability of increasingly modular and globalized satellite systems.